\documentclass[12pt]{article}
\usepackage{pdproc,epsfig} 

\begin{document}

\renewcommand{\thefootnote}{\alph{footnote}}
  
\title{A SEARCH FOR VERY HIGH ENERGY
NEUTRINOS WITH THE BAIKAL NEUTRINO TELESCOPE}

\author{ 
V.BALKANOV, L.BEZRUKOV, I.DANILCHENKO, Zh.-A.DZHILKIBAEV, G.DOMOGATSKY, \\
A.DOROSHENKO, O.GAPONENKO, A.KLABUKOV, S.KLIMUSHIN, A.KOSHECHKIN, \\
Vy.KUZNETZOV, B.LUBSANDORZHIEV, V.NETIKOV, A.PANFILOV, E.PLISKOVSKY, \\
P.POKHIL, V.POLESHUK, R.VASILJEV, V.ZHUKOV}

\address{ Institute for Nuclear Research, 60-th October Anniversary prospect,
Moscow 117312, Russia \\
 {\rm E-mail: djilkib@pcbai10.inr.ruhep.ru}}

\author{ 
N.BUDNEV, A.CHENSKY, O.GRESS, J.LJAUDENSKAITE, R.MIRGAZOV, L.PAN'KOV, \\
Yu.PARFENOV, A.PAVLOV, V.RUBZOV, Yu.SEMINEI, B.TARASHANSKY}

\address{ Irkutsk State University, Irkutsk, Russia}

\author{ 
L.KUZMICHEV, N.MOSEIKO, E.OSIPOVA, E.POPOVA, V.PROSIN, I.YASHIN}

\address{ Institute of Nuclear Physics, Moscow State University, Moscow, Russia}

\author{ 
S.FIALKOVSKY, V.KULEPOV, M.MILENIN}

\address{Nizhni Novgorod State Technical University, Nizhni Novgorod , Russia}

\author{M.ROZANOV }

\address{St.Petersburg State  Marine Technical  University, St.Petersburg, Russia}

\author{A.KLIMOV}

\address{Kurchatov Institute, Moscow, Russia}

\author{I.BELOLAPTIKOV}

\address{Joint Institute for Nuclear Research, Dubna, Russia}

\author{Ch.SPIERING, O.STREICHER, T.THON, R.WISCHNEWSKI}

\address{DESY-Zeuthen, Zeuthen, Germany}

  \centerline{\footnotesize and}

\author{D.KISS, G.TOTH}

\address{KFKI, Budapest, Hungary}

\abstract{
We present the results of a search for high energy neutrinos
with the Baikal underwater Cherenkov detector {\it NT-200.}
An upper limit on the ($\nu_e+\tilde{\nu_e}$) diffuse flux of
$E^2 \Phi_{\nu}(E)<(1.3 \div 1.9)\cdot 10^{-6}\,
\mbox{cm}^{-2}\,\mbox{s}^{-1}\,\mbox{sr}^{-1}\,\mbox{GeV}$ 
within a neutrino energy range $10^4 \div 10^7\,\mbox{GeV}$
is obtained, assuming an $E^{-2}$ behaviour of the neutrino spectrum
and flavor ratio 
$(\nu_e+\tilde{\nu_e}):(\nu_{\mu}+\tilde{\nu_{\mu}})$=1:2.}
   
\normalsize\baselineskip=15pt

\section{Introduction}
The Baikal Neutrino Telescope  is deployed in Lake 
Baikal, Siberia, 
\mbox{3.6 km} from shore at a depth of \mbox{1.1 km}. 
The optical properties of Lake Baikal deep water are characterized
by an absorption length of 20$ \div $ 25 m, a scattering length
of 20$ \div $70 m and a strongly anisotropic scattering function
$f(\theta)$ with mean cosine of scattering angle 
\mbox{$\overline{\cos}(\theta)$=0.85$\div$0.95.} 
\mbox{{\it NT-200}}, the medium-term goal of the collaboration
\cite{APP}, was put into operation on April 6th, 1998 and 
consists of 192 optical modules (OMs). 
An umbrella-like frame (see Fig.1) carries  8 strings,
each with 24 pairwise arranged OMs.
Three underwater electrical cables and one optical cable connect the
detector with the shore station. 

%%%%%%%%%%%%%%%%%%%%%%%%%%%%%%%%%%%%%%%%
\begin{figure}
%\vspace*{13pt}
\centering
%\leftline{\hfill\vbox{\hrule width 5cm height0.001pt}\hfill}
\mbox{\epsfig{figure=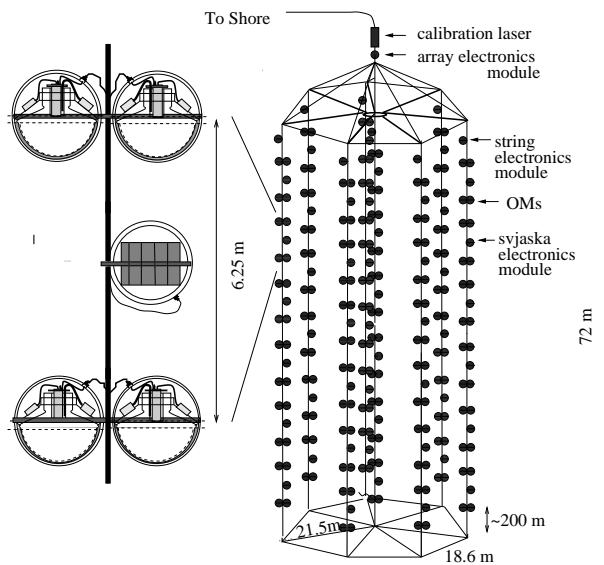,width=8.0cm}}
%\vspace*{1.4truein}		%ORIGINAL SIZE=1.6TRUEIN x 100% - 0.2TRUEIN
%\leftline{\hfill\vbox{\hrule width 5cm height0.001pt}\hfill}
\caption{Schematic view of the Baikal neutrino telescope.}
\label{fig1}
\end{figure}

%%%%%%%%%%%%%%%%%%%%%%%%%%%%%%%%%%%%%%%%
The OMs are grouped in pairs along the strings. They contain 
37-cm diameter {\it QUASAR} - photomultipliers (PMs) 
which have been developed specially for our project \cite{OM2}. 
The two PMs of a pair are switched in coincidence in order to 
suppress background from bioluminescence and PM noise. 
A pair defines a {\it channel}. 

A {\it muon-trigger}
is formed by the requirement of \mbox{$\geq $4 {\it hits}}
(with {\it hit} referring to a channel) within \mbox{500 ns}.
For  such  events, amplitude and time of all fired
channels are digitized and sent to shore. 
Full space-angular track reconstruction is possible for muon
induced events with $ \geq $6 hits at $ \geq $3 strings 
({\it off-line trigger} 6/3).
A separate {\em monopole trigger} system searches for clusters of
sequential hits in individual channels which are
characteristic for the passage of slowly moving, bright
objects like GUT monopoles.

Table 1 illustrates \mbox{\it NT-200} operation efficiency since
April 1998 till February 2001. Shown are the data taking time T, 
the fraction T/T$_{tot}$ of total detector operating time T$_{tot}$ when data
have been taken, the number of accumulated events N$_{ev}$, 
the number of events allowing full space-angular track reconstruction 
N$_{ev}$(6/3), the average fraction of working channels 
N$_{op}$/N$_{tot}$ as well as the data taking time T($>$85\%)
with $ \geq $85\%  channels in operation.  

%%%%%%%%%%%%%%%%%%%%%%%%%%%%%%%%%%%%%%%%%%%%%%%%%%%%%%%%%%%%
\begin{table}[h]
\caption{{\it NT-200} - operation efficiency }\label{tab1}
\centering
  \small
  \begin{tabular}{||c|c|c|c|c|c|c|}\hline\hline
  {} &{} &{} &{} &{} &{} &{}\\
Years & T & T/T$_{tot}$ & N$_{ev}$ & N$_{ev}$(6/3) &
N$_{op}$/N$_{tot}$ & T($>$85\%) \\
  {} & (days) &{} & (10$^6$ events) & (10$^6$ events)  &{} &(days)\\
  \hline
%  {} &{} &{} &{} &{} &{} &{}\\
  98-99 & 234 & 73\% & 167 & 57 & 71\% & 8 \\
%  {} &{} &{} &{} &{} &{} &{}\\
  \hline
%  {} &{} &{} &{} &{} &{} &{}\\
  99-00 & 236 & 75\% & 191 & 65 & 76\% & 61 \\
%  {} &{} &{} &{} &{} &{} &{}\\
  \hline
%  {} &{} &{} &{} &{} &{} &{}\\
  00-01 & 245 & 79\% & 233 & 82 & 81\% & 120 \\
%  {} &{} &{} &{} &{} &{} &{}\\
  \hline\hline
\end{tabular} 
\end{table}
%%%%%%%%%%%%%%%%%%%%%%%%%%%%%%%%%%%%%%%%%%%%%%%%%%%%%%%%%%%%

Using the data accumulated during first 234 livetime days between
April 1998 and February 1999, 35 neutrino induced upward muons
have been reconstructed \cite{B2001}. 
Although in a good agreement with MC
expectation this number is by a factor 3 lower than predicted 
for the fully operational {\it NT-200}. The reason is that,  due to unstable
operation of electronics, in average only 50 - 70 channels have taken data
 during 1998. This is in contrast to 1999 and 2000 data taking where
the stability had been improved. 
Ten events within a 30 degree half angle cone around nadir have been selected 
to set limits on the excess of the muon flux due to WIMP annihilation in 
the center of the Earth. 
Also a new limit on the flux of fast monopoles has been obtained \cite{B2001}.

Here we present preliminary results from a search for 
high energy neutrinos \mbox{(E$_{\nu}>$10 TeV)} with {\it NT-200} 
obtained from the analysis of the entire 1998 data set.
%same sample of experimental events.

\section{A search for high energy neutrinos}
The main goal of large underwater neutrino telescopes
is the search for extraterrestrial
high energy neutrinos.
Detection volume and detection area of such  detectors
depend on the transparency of the medium (water or ice) and the power of the source of
Cherenkov radiation (high energy shower or muon), and may significantly 
exceed the geometrical one. 

\subsection{Search strategy}
The used search strategy for high energy neutrinos relies
on the detection of the Cherenkov 
light emitted by the electro-magnetic and (or) hadronic
particle cascades and high energy muons
produced at the neutrino interaction
vertex in a large volume around the neutrino telescope.
Earlier, a similar strategy has been used by \mbox{DUMAND 
\cite{DUMAND}}, \mbox{AMANDA \cite{AMANDA}} and 
\mbox{BAIKAL \cite{APP3}} collaborations to obtain upper 
limits on the diffuse flux of high energy neutrinos
with relatively small detectors (SPS, AMANDA-A and \mbox{\it NT-96},
respectively). Although the limits implied by these observations
are at least one order of magnitude higher then the model independent
upper limit derived from the energy density of the diffuse
X- and gamma-radiation \cite{Ber3}, these results illustrate the 
power of used search strategy.

Neutrinos produce showers through CC-interactions

\begin{equation}
\nu_l(\bar{\nu_l}) + N \stackrel{CC}{\longrightarrow} l^-(l^+) + 
\mbox{hadrons},
\end{equation}
through NC-interactions

\begin{equation}
\nu_l(\bar{\nu_l}) + N \stackrel{NC}{\longrightarrow} 
\nu_l(\bar{\nu_l}) + \mbox{hadrons},
\end{equation}
where $l=e$, $\mu$ or $\tau$, and through resonance 
production \cite{Glash,Ber1,Gandi}

\begin{equation}
\bar{\nu_e} + e^- \rightarrow W^- \rightarrow \mbox{anything},
\end{equation}

\noindent
with the resonant neutrino energy  
$E_0=M^{2}_w/2m_e=6.3\cdot 10^6 \,$GeV 
and cross section $5.02\cdot 10^{-31}$cm$^2$.

We select events with high multiplicity of hit channels N$_{hit}$
corresponding to bright cascades. The 
volume considered for generation of cascades is essentially
{\it below} the geometrical volume of {\it NT-200}.
A cut is applied which accepts only time patterns 
corresponding to upward traveling light signals (see below). 
This cut rejects most events from brems-cascades produced by
downward going muons since the majority of muons is close to 
the vertical; they would  cross the detector and generate 
a downward time pattern. Only the fewer muons with large zenith
angles may escape detection and illuminate the array
by their proper Cherenkov radiation or via bright cascades from below the detector. 
These events then have to be rejected by a stringent multiplicity cut.

The used strategy is very effective for $\nu_e$ and $\nu_{\tau}$
detection since  the main fraction of neutrino energy 
would be transfered to electro-magnetic or/and hadronic cascades
due to CC-interactions. It is less effective for $\nu_{\mu}$ detection
since the main part of neutrino energy 
would be escaped by energetic muon from detection volume.
For $\nu_{\mu}$ search preliminary result has been presented
by the AMANDA experiment \cite{AMANDA2}.

\subsection{Data}
Within the 234 days of the detector livetime, $1.67 \cdot 10^8$ events
with $N_{hit} \ge 4$ have been selected. 
For this analysis we used events with N$_{hit}>$10. 
The time difference between any two hit channels
on the same string was required to obey the condition:

\begin{equation}
(t_i-t_j)>-10 \,\, \mbox{\rm ns}, \,\,\, (i<j),
\end{equation}
where $t_i, \, t_j$ are the arrival times at channels $i,j$ and
the numbering of channels rises from top to bottom along the string.

%%%%%%%%%%%%%%%%%%%%%%%%%%%%%%%%%%%%%%%%%%%%%%%%%%%%%%%%%%%%
\begin{figure}
%\vspace*{13pt}
\centering
\mbox{\epsfig{figure=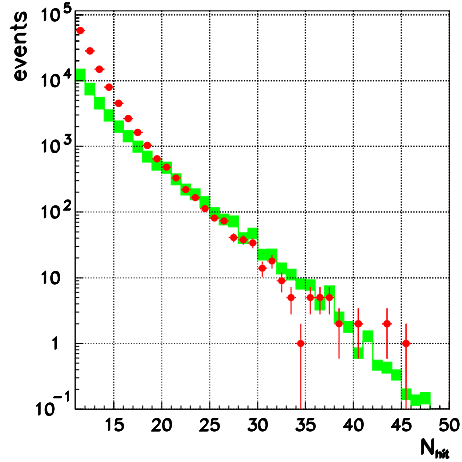,width=6.0cm}}
\caption{Distribution of hit channels multiplicity; dots - experiment,
hatched boxes - expectation from brems and hadronic showers
produced by atmospheric muons.}
\label{fig2}
\end{figure}
%%%%%%%%%%%%%%%%%%%%%%%%%%%%%%%%%%%%%%%%%%%%%%%%%%%%%%%%%%%%
Since April 1998 till February 1999, {\it NT-200} operated in 3 main
configurations with 77, 60 and 49 
working channels, respectively. 
%In Table 2 are shown the 
Data taking time T, the number of selected 
events  N$_{ev}$ which survive cut (4) and the maximum hit 
multiplicity N$_{hit}^{max}$ of these events.
are shown in Table 2.

%%%%%%%%%%%%%%%%%%%%%%%%%%%%%%%%%%%%%%%%%%%%%%%%%%%%%%%%%%%%
\begin{table}[h]
\caption{{\it NT-200} configurations in 1998.}\label{tab2}
\centering
  \small
  \begin{tabular}{||c|c|c|c|c|c|}\hline\hline
  {} &{} &{} &{} &{} &{} \\
Configuration & N$_{op}$ & T & N$_{ev}$ & N$_{hit}^{max}$ & N$_{thr}$ \\
  {} &{} &{(days)} &{} &{} &{} \\
  \hline
%  {} &{} &{} &{} &{} &{} \\
  1 & 77 & 57.9 & 63540 & 45 & 50\\
%  {} &{} &{} &{} &{} &{} \\
  \hline
%  {} &{} &{} &{} &{} &{} \\
  2 & 60 & 145.8 & 83319 & 37 & 39 \\
%  {} &{} &{} &{} &{} &{} \\
  \hline
%  {} &{} &{} &{} &{} &{} \\
  3 & 49 & 30.9 & 8719 & 24 & 26 \\
%  {} &{} &{} &{} &{} &{} \\
  \hline\hline
\end{tabular} 
\end{table}

%%%%%%%%%%%%%%%%%%%%%%%%%%%%%%%%%%%%%%%%%%%%%%%%%%%%%%%%%%%%
%A total of
%155578 events survive the selection criterion (4).
Fig.2 shows the hit multiplicity distribution
of selected (dots)
as well as the expected one from background high energy brems and
hadronic showers produced by atmospheric muons (hatched boxes).
The experimental distribution is consistent with the theoretical
expectation for N$_{hit}>$18. 
For lower N$_{hit}$ values the contribution of 
atmospheric muons close to horizon 
%nearly horizon atmospheric muons 
as well as low energy showers from $e^+e^-$ pair production
become important.
The highest multiplicity of hit channels experimentally observed is 
$N_{hit}^{max}=45$ (one event).
No statistically significant excess over expectation from atmospheric
muon induced showers has been observed for each of the 3 detector 
configurations.
The detection efficiency of {\it NT-200} for events with 
$N_{hit}>N_{hit}^{max}$ had been analysed by applying several less
stringent cuts. It was shown that the experimental $N_{hit}$ distributions
are consistent with expected ones from atmospheric muons.

Since no events with $N_{hit}>N_{hit}^{max}$ 
are found in our data we can derive upper limits on the flux of 
high energy neutrinos which would produce events with 

\begin{equation}
N_{hit}>N_{thr},
\end{equation}
where the chosen values of N$_{thr}$ for the 3 
detector configurations are given in \mbox{Table 2}. 

\subsection{MC-simulations}
Given an isotropic diffuse high energy neutrino flux with power law energy 
spectrum with spectral index $\gamma$, the number of expected events during
observation time T reads 

%%%%%%%%%%%%%%%%%%%%%%%%%%%%%%%%%%%%%%%%%%%%%%%%%%%%%%%%%%%% 
\begin{figure}
%\vspace{-2.0cm}
\centering
%\mbox{\epsfig{figure=n_gamma.eps,width=6.5cm}}
\mbox{\epsfig{figure=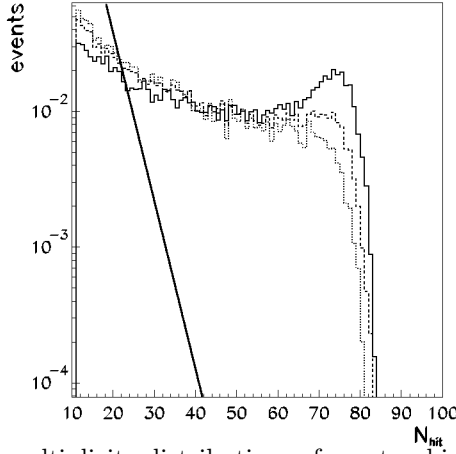,width=6.0cm}}
\vspace{-0.5cm}
\caption{The normalized hit multiplicity distributions of events
which would be produced by the $\nu_e$ fluxes and survive the 
selection criterion (4). Solid, dashed and dotted curves correspond
to $\gamma$=1.5, 2, 2.5, respectively. 
Also shown is 
the normalized N$_{hit}$ distribution
of events from atmospheric muon induced showers (thick line).
}
\label{fig3}
\end{figure}
%%%%%%%%%%%%%%%%%%%%%%%%%%%%%%%%%%%%%%%%%%%%%%%%%%%%%%%%%%%%

\begin{equation}
N_{\nu}=\frac{A_{\nu}T}{4\pi} \int d\Omega \int dEV_{eff}(\Omega,E)\sum_k 
\int
dE_{\nu} E_{\nu}^{- \gamma} N_{A} \rho_{H_2O} \frac{d\sigma_{\nu k}}{dE}
\exp(-l(\Omega)/l_{tot})
\end{equation}
where $E_{\nu}$ is the neutrino energy, $E$ - the energy 
transferred to a shower, $A_{\nu}$ - normalization
coefficient of neutrino flux and $V_{eff}(\Omega , E)$ - detection volume
\footnote{In a case of $\nu_{\mu}$ CC- interaction in water the detector response to
hadronic shower as well as to the high energy muon
has been taken into account.}.
The index $\nu$ indicates neutrino types
($\nu=\nu_{\mu},\tilde{\nu_{\mu}},\nu_e, \tilde{\nu_e}$) and
$k$ indicates the summation over CC and NC interactions.
$N_{A}$ is the Avogadro number. Cross sections \cite{Ber2,Gandi} $d\sigma_{\nu k}/dE$ 
correspond to processes (1) and (2). The neutrino absorption in the Earth 
has been taken into account with a suppression factor 
$\exp(-l(\Omega)/l_{tot})$, where $l(\Omega)$ is the neutrino 
path length through the Earth in direction $\Omega$ and 
$l^{-1}_{tot}=N_A \, \rho_{Earth} \, (\sigma_{CC}+\sigma_{NC})$ 
according \mbox{to \cite{Ber2,Gandi}.}
%%%%%%%%%%%%%%%%%%%%%%%%%%%%%%%%%%%%%%%%%%%%%%%%%%%%%%%%%%%% 
%\vspace{-1.5cm}
\begin{figure}
%\vspace{0.3cm}
\centering
%\mbox{\epsfig{figure=frac_90.eps,width=6.5cm}}
\mbox{\epsfig{figure=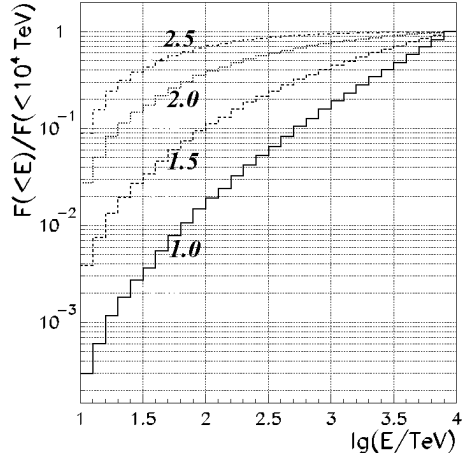,width=6.0cm}}
\caption{The fractions of expected events induced by diffuse
$\nu_e$ fluxes with spectral indexes $\gamma$=1, 1.5, 2, 2.5
within energy range 10 TeV$<$E$_{\nu}<$E.} 
%\vspace{-0.5cm}
\label{fig4}
\end{figure}
%%%%%%%%%%%%%%%%%%%%%%%%%%%%%%%%%%%%%%%%%%%%%%%%%%%%%%%%%%%%

In Fig.3 we show the normalized 
predictions of N$_{hit}$ distributions of events 
which survive cut (4) and would be induced by electron 
neutrino fluxes with \mbox{$\gamma$=1.5, 2, 2.5.} 
The normalized N$_{hit}$ distribution of events from 
atmospheric muon induced showers, which has strongly 
steeper behaviour, is also presented.

The neutrino detection energy range of {\it NT-200} which contains, 
for instance, 90\% of 
expected events, depends on the value of spectral index $\gamma$.
Normalized energy distributions of expected events induced by 
electron neutrinos within energy range 
10 TeV $\div$ E (E$<$10$^4$ TeV)
and different $\gamma$ are presented in Fig.4.
Assuming $\gamma=$2 as typically expected for Fermi acceleration, 
90\% of expected events would be produced by  neutrinos
from the energy range $20 \div 10^4$ TeV with the mean energy 
around 200 TeV.

%%%%%%%%%%%%%%%%%%%%%%%%%%%%%%%%%%%%%%%%%%%%%%%%%%%%%%%%%%%%
\begin{figure}
\vspace{-0.8cm}
\centering
\mbox{\epsfig{figure=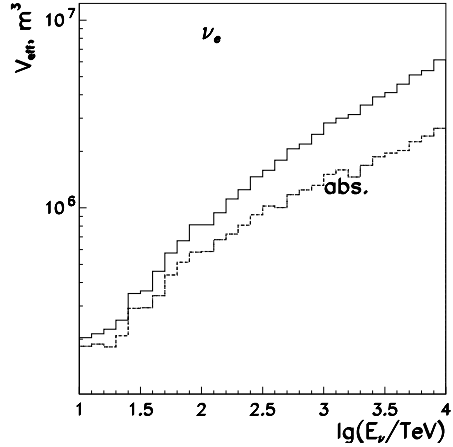,width=6.5cm}}
\caption{Detection volume of $\nu_e$ produced events which
survive cuts (4)-(5) (upper curve). 
%Also shown is the
The detection volume folded with the neutrino absorption probability
in the Earth (lower curve) is also shown.}
%\vspace{-1.5cm}
\label{fig5}
\end{figure}
%%%%%%%%%%%%%%%%%%%%%%%%%%%%%%%%%%%%%%%%%%%%%%%%%%%%%%%%%%%%
The detection volume for neutrino produced events which fulfill
conditions (4)-(5) was calculated as a function of neutrino energy
and zenith angle $\theta$. The energy dependence of the detection 
volume for isotropic $\nu_e$ flux with $\gamma =$2 is shown in Fig.5.
Also shown is the detection volume folded with the neutrino
absorption probability in the Earth. The value of $V_{eff}$ rises 
from 2$\cdot$10$^5$ m$^3$ for 10 TeV up to 6$\cdot$10$^6$ m$^3$    
for $10^4$ TeV and significantly exceeds the geometrical volume
$V_{g} \approx$ 10$^5$ m$^3$ of {\it NT-200}. This is
due to the low light scattering and the preserved light fronts
from Cherenkov waves originating far outside the geometrical
volume. 
%%%%%%%%%%%%%%%%%%%%%%%%%%%%%%%%%%%%%%%%%%%%%%%%%%%%%%%%%%%%%%
Although the detection volume has been calculated without
taking into account light scattering in the water, estimations
show that a scattering with
L$_s$=20 m and \mbox{$\overline{\cos}(\theta)$=0.88} 
(conservative values for Lake Baikal water) would cause
$\leq$30\% decrease of V$_{eff}$ for E$_{\nu} \leq$10$^3$TeV.

%%%%%%%%%%%%%%%%%%%%%%%%%%%%%%%%%%%%%%%%%%%%%%%%%%%%%%%%%%%%

\begin{figure}
%\vspace{0.5cm}
\centering
\mbox{\epsfig{figure=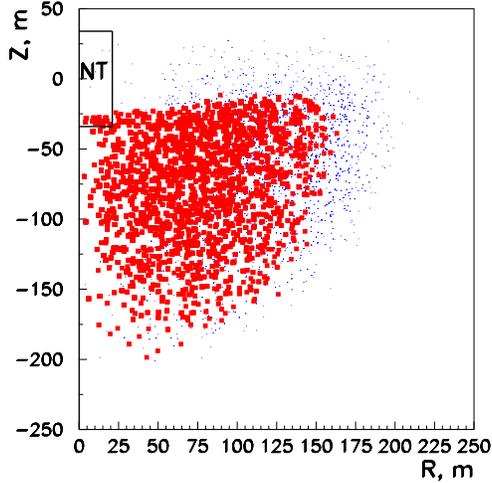,width=7.0cm}}
\caption{Coordinates of $\nu_e$ interaction vertex
for events which fulfill conditions (4) (dots) and 
(4)-(5) (rectangles).}
%\vspace{-0.5cm}
\label{fig6}
\end{figure}
%%%%%%%%%%%%%%%%%%%%%%%%%%%%%%%%%%%%%%%%%%%%%%%%%%%%%%%%%%%%

Fig.6  illustrates the difference between $V_{eff}$ and $V_{g}$. 
Shown here are the coordinates of neutrino interaction vertex 
for events which survive cuts (4) (dots) and 
(4)-(5) (rectangles). 
 
\subsection{The limits on the high energy neutrino fluxes}
The shape of the neutrino spectrum was assumed to behave like 
$E^{-2}$ and flavor ratio 
$(\nu_e+\tilde{\nu_e}):(\nu_{\mu}+\tilde{\nu_{\mu}})=1:2$  
due to photo-meson production of $\pi^+$ followed by
the decay $\pi^+ \rightarrow \mu^+ + \nu_{\mu} \rightarrow 
e^+ + \nu_e + \bar{\nu_{\mu}} +\nu_{\mu}$  
for extraterrestrial sources. 

Comparing the expected number of events fulfilling (4)-(5)
with the upper limit on the actual number of events, 2.4 for 90\% C.L.
\cite{FELDMAN}
we obtain the upper limit on the diffuse ($\nu_e+\tilde{\nu_e}$) flux.
The combined upper limit obtained with Baikal neutrino telescopes 
{\it NT-200} (234 days) and {\it NT-96 \cite{APP3}} (70 days) 
is:

\begin{equation}
\Phi_{(\nu_e+\tilde{\nu_e})}E^2<(1.3 \div 1.9)\cdot10^{-6} 
\mbox{cm}^{-2}\mbox{s}^{-1}\mbox{sr}^{-1}\mbox{GeV},
\end{equation}
where the upper value refers to the conservative limit on light
scattering in the Baikal water.

%%%%%%%%%%%%%%%%%%%%%%%%%%%%%%%%%%%%%%%%%%%%%%%%%%%%%%%%%%%%
\begin{figure}
\vspace{-1.5cm}
\centering
\mbox{\epsfig{figure=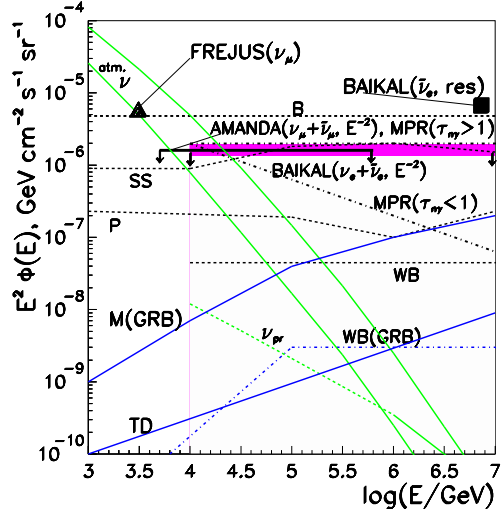,width=6.5cm}}
\caption{The upper limits on the diffuse neutrino fluxes, obtained
by BAIKAL
(this work: for E$^{-2}$ flux as well as for a model independent $\tilde{\nu_e}$
flux at resonant energy 6.3$\cdot$10$^6$GeV (rectangle)),
AMANDA
%\cite{AMANDA2} 
and FREJUS
%\cite{FREJUS} 
(triangle) experiments as well as the atmospheric conventional 
neutrino fluxes 
%\cite{VOL} 
from horizontal and vertical directions (upper and lower curves, respectively) 
and atmospheric prompt neutrino flux 
%\cite{PROMT} 
(curve labeled $\nu_{pr}$). 
The limits and predictions from different models of high energy neutrino
sources are also shown. Curves labeled 'B', 'MPR' and 'WB' represent the model independent
limit derived by Berezinsky 
%\cite{Ber3} 
and  the upper bounds derived by Mannheim et al. 
%\cite{P98} 
as well as by Waxman and Bahcall. 
%\cite{WB1}.
Curves labeled as 'SS' and 'P' show predicted fluxes from quasar cores 
%and blazar jets derived by Stecker and Salamon 
%\cite{SS} 
and Protheroe 
%\cite{P} 
respectively.
Curves labeled 'M(GRB)' and 'WB(GRB)' represent predicted fluxes
from GRBs derived by Mannheim 
%\cite{MANNHEIM} 
and Waxman and Bahcall.
% \cite{WB2}. 
Curve labeled 'TD' - prediction for neutrino flux from topological
defects due to specific top-down scenario BHS1.
% \cite{BHS1}.
}
\label{fig7}
\end{figure}
%%%%%%%%%%%%%%%%%%%%%%%%%%%%%%%%%%%%%%%%%%%%%%%%%%%%%%%%%%%%

Fig.7 shows the upper limits on the isotropic diffuse high energy neutrino
fluxes obtained by BAIKAL (this work), 
AMANDA \cite{AMANDA2}
and FREJUS \cite{FREJUS} (triangle) experiments 
as well as the atmospheric conventional neutrino \mbox{fluxes \cite{VOL}} 
from horizontal and vertical directions (upper and lower curves,
respectively) and atmospheric prompt neutrino flux \cite{PROMT} 
(curve labeled $\nu_{pr}$). 
Also shown is the model  
independent upper limit on the diffuse high energy neutrino flux
obtained by V.Berezinsky \cite{Ber3} (curve labeled 'B')
with the energy density of the diffuse X- and gamma-radiation 
$\omega_x \leq 2 \cdot 10^{-6}$ eV cm$^{-3}$ (as follows from
EGRET data \cite{EGRET}),  and
predictions for diffuse neutrino fluxes
from Stecker and Salamon model \cite{SS} 
(the sum of the quasar core and blazar jet contribution, 
curve labeled 'SS') and 
Protheroe \mbox{model \cite{P}}   (from equal contribution
from $pp$ and $p \gamma$ interactions in AGN jets, curve labeled 'P').
Curves labeled 'MPR' and 'WB' show the upper bounds obtained by Mannheim, Protheroe
and Rachen \cite{P98} for optically thick ($\tau_{n\gamma}>1$) and
optically thin ($\tau_{n\gamma}<1$) sources as well as the upper bound obtained
by Waxman and Bahcall \cite{WB1} for optically thin sources respectively.
Curves labeled 'M(GRB)' and 'WB(GRB)' present the upper bounds for diffuse
neutrino flux from GRBs derived by Mannheim \cite{MANNHEIM} and
Waxman and Bahcall \cite{WB2}. 
Curve labeled 'TD' shows prediction for neutrino flux from topological
defects due to specific top-down scenario BHS1 \cite{BHS1}.

For the resonant process (3) the event number is given by:

\begin{equation}
N_{\bar{\nu_e}}=T \int d\Omega
\int dEV_{eff}(\Omega,E) 
\int\limits_{(M_w-2\Gamma_w)^2/2m_e}^{(M_w+2\Gamma_w)^2/2m_e}
dE_{\nu}\Phi_{\bar{\nu_e}}(E_{\nu}) 
\frac{10}{18}N_{A} \rho_{H_2O} \frac{d\sigma_{\bar{\nu_e},e}}{dE}
\end{equation}

$$
M_w=80.22 \mbox{GeV}, \, \, \, \Gamma_w=2.08 \mbox{GeV}.
$$

Our combined 90\% C.L. limit obtained with {\it NT-200} and 
{\it NT-96 \cite{APP3}} at the W - resonance energy is:

\begin{equation}
\frac{d\Phi_{\bar{\nu}}}{dE_{\bar{\nu}}} \leq (1.4 \div 1.9) \times 
10^{-19} 
\mbox{cm}^{-2}\mbox{s}^{-1}\mbox{sr}^{-1}\mbox{GeV}^{-1}.
\end{equation}
and is also depicted in Fig.7 (rectangle).

\section{Conclusion}
The deep underwater neutrino telescope {\it NT-200} in Lake Baikal is 
taking data since April 1998. Due to the high water transparency and 
low light scattering, the detection volume of {\it NT-200} for high 
energy $\nu_e$ and $\nu_{\tau}$ detection is several megatons and
exceeds the geometrical volume by factor of about 50 for highest
energies.  
This permits a search for diffuse neutrino fluxes from extraterrestrial 
sources on the level of theoretical predictions. The upper limits (7), (9) 
obtained for the diffuse E$^{-2}$ ($\nu_e+\tilde{\nu_e}$) flux and the model 
independent $\tilde{\nu_e}$ flux at resonant energy 6.3$\cdot$10$^6$GeV 
are the most stringent ones at present. We expect that the analysis of 3 
years data taken with {\it NT-200} would allow us to 
%investigate high energy neutrinos on a level 
reach a sensitivity of
$\Phi_{\nu}E^2 \approx 6\cdot10^{-7}$cm$^{-2}$s$^{-1}$sr$^{-1}$GeV.

\section{Acknowledgements}
This work was supported by the Russian Ministry of Research 
(contract \mbox{\sf 102-11(00)-p}), the German 
Ministry of Education and Research and the Russian Fund of Basic 
Research (grants  \mbox{\sf 99-02-18373a}, 
\mbox{\sf 01-02-31013} and \mbox{\sf 00-15-96794}),
and by the Russian Federal Program ``Integration'' (project no. 346).


\begin{thebibliography}{99}
\bibitem{APP} I.A.Belolaptikov {\it et al.}, 
{\it Astropart. Phys.} {\bf 7} (1997) 263. 

\bibitem{OM2} R.I.Bagduev {\it et al.,} {\it Nucl. Instr. Meth.} 
{\bf A420} (1999) 138.

\bibitem{B2001} V.A.Balkanov {\it et al.}, 
{\it Nucl. Phys. Proc. Suppl.} {\bf 91} (2001) 438. 

\bibitem{DUMAND} J.W.Bolesta {\it et al.}, 
{\it Proc. 25-th ICRC} Durban--South Africa, {\bf 7} (1997) 29.

\bibitem{AMANDA} R.Porrata {\it et al.}, {\it Proc. 25-th ICRC} 
Durban--South Africa, {\bf 7} (1997) 9.

\bibitem{APP3} V.A.Balkanov {\it et al.}, 
{\it Astropart. Phys.} {\bf 14} (2000) 61. 

\bibitem{AMANDA2} E.Andres {\it et al.}, {\it Nucl. Phys. Proc. Suppl.} 
{\bf 91} (2001) 423.

\bibitem{Ber3} V.S.Berezinsky {\it et al.,}
{\it Astrophysics of Cosmic Rays,} North Holland,
Amsterdam (1990).

\bibitem{Glash} S.L.Glashow, {\it Phys. Rev.} {\bf 118} (1960) 316.

\bibitem{Ber1} V.S.Berezinsky and A.Z.Gazizov,
{\it JETP Lett.} {\bf 25} (1977) 254.

\bibitem{Ber2} V.S.Berezinsky {\it et al.,}
{\it Sov. J. Nucl. Phys.} {\bf 43} (1986) 406.

\bibitem{Gandi} R.Gandhi {\it et al.},  
{\it Astropart. Phys.} {\bf 5} (1996) 81.

\bibitem{FELDMAN} G.Feldman and R.Cousins,
{\it Phys. Rev.} {\bf D57} (1998) 3873.

%\bibitem{APP2} I.A.Belolaptikov {\it et al.}, 
%{\it Astropart. Phys.} {\bf 12} (1999) 75. 

%\bibitem{JF} V.A.Balkanov {\it et al.}, {\it Physics of Atomic Nuclei} 
%{\bf 62} (1999) 949.

%\bibitem{Project}
%BAIKAL Collaboration, {\it The Baikal Neutrino Telescope NT-200, 
%BAIKAL 92-03}, ed. by I.A.Sokalski and Ch.Spiering (1992).

%\bibitem{BELEN} S.Z.Belenkij, {\it Cascade processes in
%cosmic rays}, Gostehizdat (Moscow) (1948).

%\bibitem{ALV1} J.Alvarez-Muniz and E.Zas, {\it Physics Letters}
%{\bf B411} (1997) 218.

%\bibitem{ALV2} J.Alvarez-Muniz and E.Zas, {\it astro-ph/9906347}.

%\bibitem{EAS} M.Aglietta {\it et al.}, {\it Physics Letters} 
%{\bf B333} (1994) 555.


\bibitem{FREJUS} W.Rhode {\it et al.}, {\it Astropart. Phys.} 
{\bf 4} (1994) 217.

\bibitem{VOL} L.Volkova,  {\it Yad.Fiz.} {\bf 31} (1980) 1510
({\it Sov. J. Nucl. Phys.} {\bf 31} (1980) 784).

%\bibitem{LIP} P.Lipari,  {\it Astropart. Phys.} {\bf 1} (1993) 195.

\bibitem{PROMT} M.Thunman, G.Ingelman and P.Gondolo,
{\it Astropart. Phys.} {\bf 5} (1996) 309.


\bibitem{EGRET} P.Sreekumar  {\it et al.} 
(EGRET Collaboration), {\it Ap. J.} {\bf 494} (1998) 523.

\bibitem{SS} F.W.Stecker and M.H.Salamon, 
{\it Astro-ph/9501064.}

\bibitem{P} R.Protheroe, in
{\it Accretion Phenomena and Related Outflows,}
Vol. 163 of {\it IAU Colloquium,} edited by
D.Wickramasinghe, G.Bicknell and L.Ferrario
(The Astron. Soc. of the Pacific, 1997), pp.585-588,
{\it astro-ph/9809144.}

\bibitem{P98} K.Mannheim, R.J.Protheroe and J.P.Rachen,   
{\it astro-ph/9812398}. 

\bibitem{WB1} E.Waxman and J.Bahcall,
{\it Phys. Rev.} {\bf D 59} (1999) 023002.

\bibitem{MANNHEIM} K.Mannheim, 
{\it astro-ph/0010353}. 

\bibitem{WB2} E.Waxman and J.Bahcall,
{\it Phys. Rev. Lett.} {\bf 78} (1997) 2292.

\bibitem{BHS1} P.Bhattacharjee, C.Hill and D.Schramm,
{\it Phys. Rev. Lett.} {\bf 69} (1992) 567;
G.Sigl, {\it astro-ph/0008364}. 


\end{thebibliography}
\end{document}